\documentclass[runningheads]{llncs}
\usepackage[T1]{fontenc}
\usepackage{graphicx,verbatim}
\usepackage{color}
\usepackage{amssymb}
\usepackage{amsmath} % Ensure amsmath is loaded
\usepackage[dvipsnames]{xcolor}
\usepackage{amsmath}
\usepackage{bm}
\usepackage{soul}
\usepackage{multirow}
\usepackage{pifont}% http://ctan.org/pkg/pifont
\newcommand{\cmark}{\ding{51}}%
\newcommand{\xmark}{\ding{55}}%
\DeclareMathOperator{\ReLU}{ReLU}
\DeclareMathOperator{\vecop}{vec}
\usepackage{hyperref}

\begin{document}
%
% COMMENT: Important to metion US
\title{Fusing Radiomic Features with Deep Representations for Gestational Age Estimation in Fetal Ultrasound Images}
% COMMENT: This running title is just for the page header
\titlerunning{Fusing Radiomic Features and DL Representations for GA Estimation}

%
% \begin{comment}  %% Removed for anonymized MICCAI 2025 submission
\author{
Fangyijie Wang\inst{1,2} \and % index{Wang, Fangyijie}
Yuan Liang\inst{1,4} \and % index{Liang, Yuan}
Sourav Bhattacharjee\inst{3} \and % index{Bhattacharjee, Sourav}
Abey Campbell\inst{4} \and % index{Campbell, Abey}
Kathleen M. Curran\inst{1,2} \and % index{Curran, Kathleen M.}
Gu\'enol\'e Silvestre\inst{1,4} % index{Silvestre, Gu\'enol\'e}
}
\authorrunning{F. Wang et al.}
\institute{
Research Ireland Centre for Research Training in Machine Learning  \and
School of Medicine, University College Dublin, Dublin, Ireland \and
School of Veterinary Medicine, University College Dublin, Dublin, Ireland \and
School of Computer Science, University College Dublin, Dublin, Ireland 
\email{fangyijie.wang@ucdconnect.ie} \\
}

% \end{comment}

% \author{Anonymized Authors}  %% Added for anonymized MICCAI 2025 submission
% \authorrunning{Anonymized Author et al.}
% \institute{Anonymized Affiliations \\
%     \email{email@anonymized.com}}

\maketitle              % typeset the header of the contribution
\begin{abstract}
Accurate gestational age (GA) estimation, ideally through fetal ultrasound measurement, is a crucial aspect of providing excellent antenatal care. However, deriving GA from manual fetal biometric measurements depends on the operator and is time-consuming. Hence, automatic computer-assisted methods are demanded in clinical practice. In this paper, we present a novel feature fusion framework to estimate GA using fetal ultrasound images without any measurement information. 
% Although deep learning models demonstrate outstanding performance on feature extraction, they still fall short in estimating GA. 
We adopt a deep learning model to extract deep representations from ultrasound images. We extract radiomic features to reveal patterns and characteristics of fetal brain growth. To harness the interpretability of radiomics in medical imaging analysis, we estimate GA by fusing radiomic features and deep representations.
Our framework estimates GA with a mean absolute error of 8.0 days across three trimesters, outperforming current machine learning-based methods at these gestational ages. Experimental results demonstrate the robustness of our framework across different populations in diverse geographical regions. Our code is publicly available on \href{https://github.com/13204942/RadiomicsImageFusion_FetalUS}{GitHub}.


\end{abstract}
\section{Introduction}
Gestational age (GA) is the common term used during pregnancy to know the duration of gestation. This estimation is based on the assumption of an average fetal size at each stage of gestation. A normal pregnancy can range from 37 to 42 weeks \cite{MedlinePlus}. 
% Traditionally, GA is estimated using the first day of the woman's last menstrual period (LMP).
Accurate dating of pregnancy is crucial for effective pregnancy management throughout all stages, from the first trimester to delivery \cite{Papageorghiou:2016}. It is essential for identifying premature labor and postdated deliveries \cite{Papageorghiou:2016}. Therefore, precision estimation of GA enables the accurate scheduling of antenatal care for women, guides obstetric management decisions, and supports the proper interpretation of fetal growth assessments \cite{Sinha:2020}.

% The traditional estimation of GA is using the first day of the last menstrual period (LMP). This method is based on the assumption that ovulation takes place on the 14th day of the menstrual cycle \cite{MedlinePlus}. 
However, in many settings, women attend their antenatal care late in pregnancy or even upon delivery, which presents challenges in accurately estimating GA \cite{Papageorghiou:2016}. Ultrasound imaging is widely used for screening and monitoring of pregnant women, as well as fetal growth \cite{Loughna:2009,Salomon:2011}. In the screening examination, biometric measurements of the fetus including the crown-rump length (CRL), head circumference (HC), abdominal circumference (AC), and femur length (FL), are commonly calculated to estimate the GA \cite{Heuvel:2018_b,Sinha:2020}. 

Recent advances in deep learning (DL) algorithms for fetal ultrasound image analysis have shown that significant potential of deep representations in the accuracy of diagnosis and biometric estimation \cite{Fiorentino:2023,Naz:2025}.
% The recent advancements in deep learning (DL) algorithms have significantly improved the accuracy of GA estimation \cite{Zhang:2020,Lee:2023,Fiorentino:2023,Lee:2023_b,Guo:2024}.
Several studies have employed regression algorithms to estimate GA from ultrasound images \cite{Lee:2023,Lee:2023_b}. 
These methodologies primarily rely on deep representations extracted by DL models, except \cite{Guo:2024}, which provides an automated multimodal pipeline to estimate biometric parameters and GA at the same time. 
DL models exhibit exceptional performance in feature extraction; however, their lack of interpretability has restricted the development of clinical applications in estimating GA.
% Recent advancements in deep learning (DL) algorithms for fetal ultrasound image analysis have shown that valid deep representations can benefit the accuracy of diagnosis and biometric estimation \cite{Keerthi:2024,Guo:2024,Cho:2024}.
%\cite{Keerthi:2024,Guo:2024,Venturini:2025,Plotka:2025}. 
Prior studies have shown that ultrasound images obtained during pregnancy can offer valuable insights into fetal brain maturation, assisting in estimating GA for screening purposes \cite{Namburete:2017,Welp:2020}. 
Recent research \cite{Zilka:2024} demonstrates that radiomic features capture physics-related characteristics, making them intrinsically interpretable. However, there is a gap in the existing research as no study has yet explored to combine radiomic features with DL techniques for fetal ultrasound imaging analysis. 
To our knowledge, this study represents the pioneering effort in creating an interpretable pathway that combines radiomic features and deep representations to estimate GA in fetal ultrasound images.

In this paper, we present a novel feature fusion framework that combines radiomic features and deep representations of the fetal head extracted by a convolutional neural network (CNN) to improve the predictive accuracy of GA.
We evaluate the performance of our framework using two different fetal head ultrasound datasets and present quantitative results obtained from machine learning (ML) and DL models for comparison.
The main contributions of our work: (i) we propose the first-ever feature fusion framework that combines fetal ultrasound images and radiomics to estimate GA; (ii) we present a plug-and-play cross-attention module for the effective fusion of radiomic features and deep representations; (iii) finally, comprehensive experiments demonstrate the effectiveness of our fusion module in estimating GA on ultrasound data.

\begin{figure}[htbp]
\centering
\includegraphics[width=\textwidth]{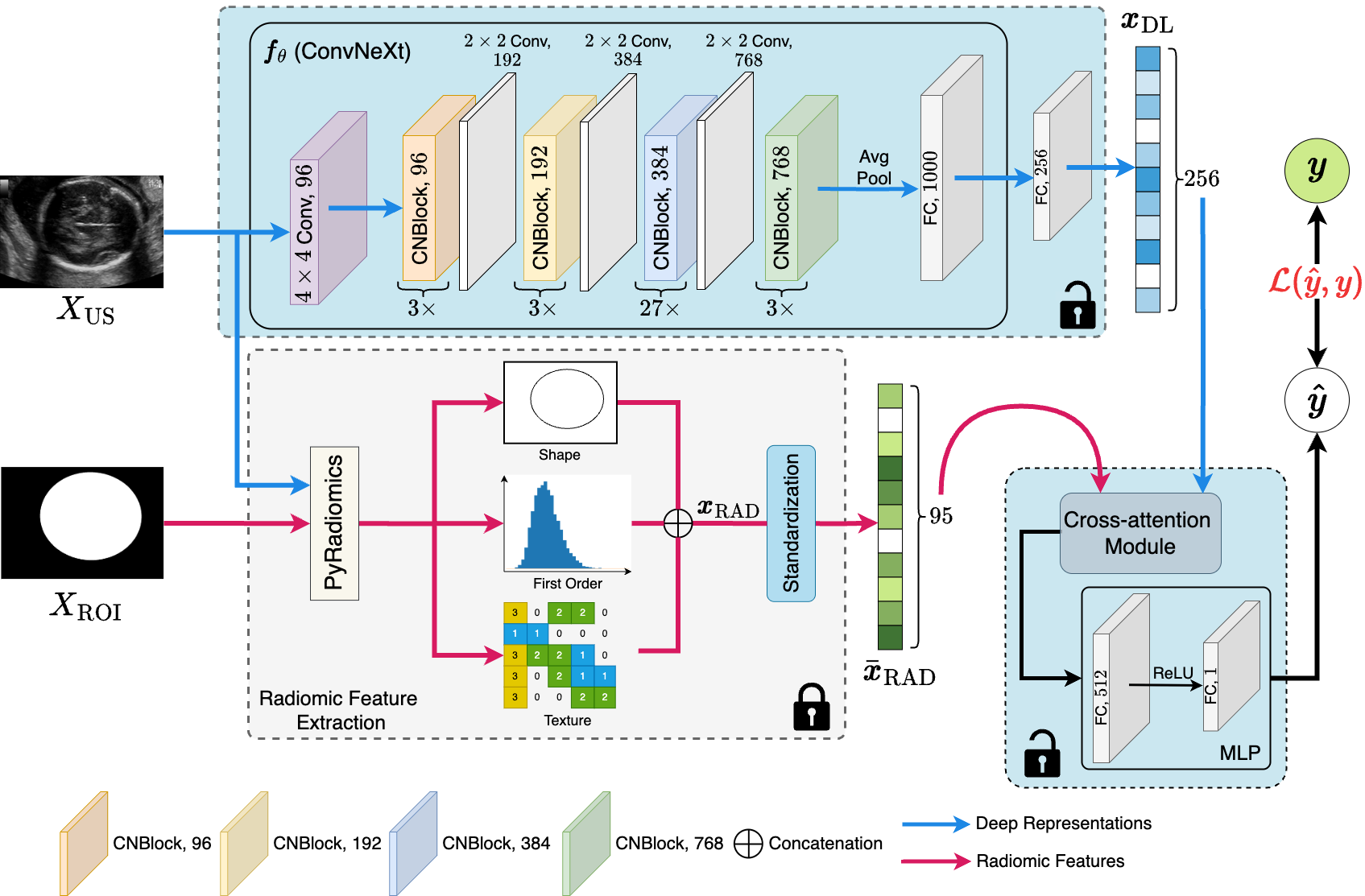}
\caption{Proposed fusion framework for GA estimation from fetal ultrasound images. \textbf{\color{SkyBlue} Blue Box}: Trainable parameters (DL model, cross-attention, MLP). \textbf{\color{gray} Gray Box}: Non-trainable parameters (Radiomics).}\label{fusion_network}
\end{figure}

\section{Methodology}

\subsection{Overview}
An overview of our proposed fusion framework for estimating GA with radiomic features and deep representations is illustrated in Fig.~\ref{fusion_network}. The framework contains three core components: (1) a pre-trained DL model, ConvNeXt \cite{Liu:2022}, is used for deep representation learning; (2) a Python pipeline extracts radiomic features; (3) a cross-attention module for feature fusion.

\subsection{Deep Representation Learning}

\begin{figure}[htbp]
\centering
\includegraphics[width=.9\textwidth]{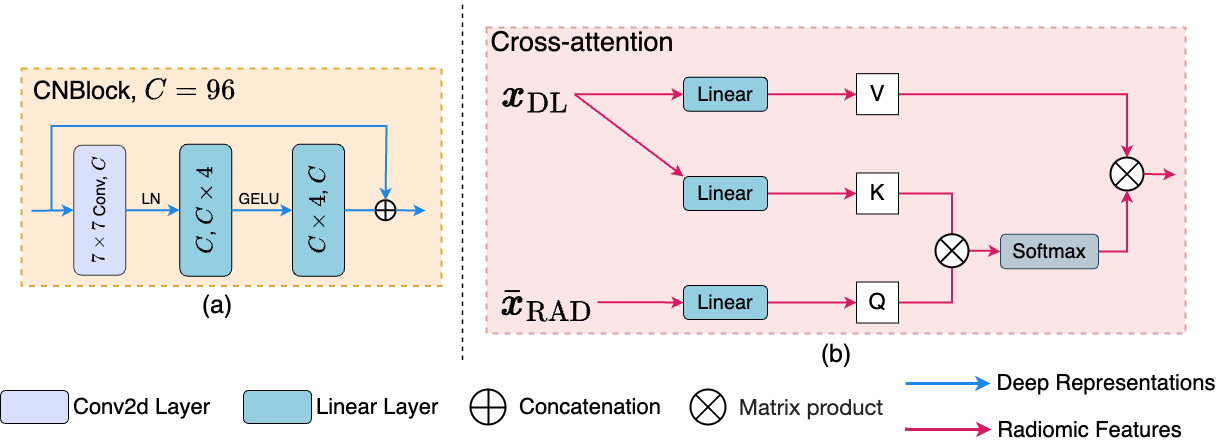}
\caption{Visualization of CNBlock and Cross-attention module. \textbf{(a)} is an example of CNBlock with channel size 96; \textbf{(b)} represents the cross-attention module for integrating radiomic features and deep representations. LN: Layer Normalization} \label{modules}
\end{figure}

In our proposed framework, we employ the CNN topology to construct the DL model $f_{\theta}$ to learn deep representations that reflect high-level semantic information of ultrasound images, where, $\theta$ represents the learnable parameters of the model. 
% COMMENT: repeat In our comprehensive experiments, we employ CNNs and Transformers to construct $f_{\theta}$. 
Specifically, we incorporate models such as ResNet18 \cite{He:2016}, SwinTransformer \cite{Liu:2022b}, EfficientNet \cite{Tan:2021}, MobileNet V3 \cite{Howard:2019}, ViT \cite{Dosovitskiy:2020_b}, MaxViT \cite{Tu:2022}, and ConvNeXt \cite{Liu:2022} for their proven effectiveness in image classification tasks. %The technical details of these DL architectures are documented in prior research \cite{He:2016,Howard:2019,Dosovitskiy:2020,Tan:2021,Liu:2022b,Tu:2022,Liu:2022}. 
The DL model $ f_{\theta} $ in our proposed framework adopts the ConvNeXt stack architecture with four layers. Each layer consists of a CNBlock followed by a convolution layer. The CNBlock layers have channel size of 96, 192, 384, and 768, respectively. The detailed illustration of the CNBlock structure is shown in Fig.~\ref{modules}(a).

% \hl{COMMENT: superscript dimensions for X is not very consistent. Maybe drop and instead say $X\in\mathbb{R}^{3\times n_h\times n_w}$. Keep consistent notation for all dimensions, maybe something like lowercase italic $n_{whatever}$. So when you encounter $n$ with subscript you know that it is a dimension of some kind. A}

% {\color{blue} When working with a grayscale fetal ultrasound image $X_\text{US}$, the initial step involves converting $X_\text{US}$ into a 3-channel image denoted as $X_\text{US}\in\mathbb{R}^{3\times n_\text{h}\times n_\text{w}}$, where $n_\text{h}$ and $n_\text{w}$ represent the height and width of the image, respectively. This conversion is essential to facilitate the fine-tuning of pre-trained DL models designed to process 3-channel input images.}
The input $X_\text{US}\in\mathbb{R}^{3\times n_\text{h}\times n_\text{w}}$ is processed by the model $f_{\theta}$, producing a deep representation vector that is subsequently fed to the $\ReLU$ activation and linearly combined to output $\bm{x}_\text{DL}\in\mathbb{R}^{512}$. %, as follows:
% \begin{align}
% \label{deepmodel_eq}
% \bm{x}_\text{DL} & = W_{\text{DL}} \cdot f_{\theta}(X_\text{US}) + b_{\text{DL}}
% \end{align}
% The input ultrasound data $X_\text{US}$ has dimensions of $3 \times n_\text{h} \times n_\text{w}$, where $n_\text{h}$ and $n_\text{w}$ represent the height and width of the image, respectively.
Subsequently, $\bm{x}_\text{DL}$ is fused with radiomic features $\bar{\bm x}_{\text{RAD}}$ within the cross-attention module.

\subsection{Radiomic Features}

% \hl{COMMENT: in order to get rid of the ugly superscript, how about calling $X_\text{US}$ the $3\times n_h\times n_w$ input and $X_\text{ROI}$. Just provide dimensions in text.}

Our dataset comprises fetal head ultrasound images and corresponding region of interest (ROI) masks, denoted as $X_\text{US}$ and $X_\text{ROI}$, respectively. %{\color{blue} The ROI encompasses the full extent of the fetal head, as annotated by sonographers.} 
The input $X_\text{ROI}$ has dimensions of $1 \times n_\text{h} \times n_\text{w}$, where $n_\text{h}$ and $n_\text{w}$ correspond to the height and width of ROI image. 
The process of extracting radiomic features utilizes each pair of $X_\text{US}$ and $X_\text{ROI}$ as input. In this study, a total of 95 radiomic features $\bm{x}_{\text{RAD}} \in \mathbb{R}^{95}$, which capture surface-level details of the fetal head region, are extracted with the publicly accessible Python package, \verb|pyradiomics|\footnote{https://pyradiomics.re adthedocs.io/en/latest/}. These extracted features are categorized into shape, statistical (first-order), and texture features \cite{Fan:2021,Aerts:2014,Griethuysen:2017}.

% The size of features in six categories is 9, 18, 22, 16, 16 and 14. 
The radiomic features can be categorized into six groups: shape features (Shape2D), First-order features, texture features consist of gray level co-occurrence matrix (GLCM), gray level run length matrix (GLRLM), gray level size zone matrix (GLSZM), and gray level dependence matrix (GLDM) features. Before the feature fusion step, we standardize radiomic features $\bm x_{\text{RAD}} $ by removing the mean and scaling to unit variance to obtain $\bar{\bm x}_{\text{RAD}} \in \mathbb{R}^{95}$. Subsequently, these standardized features $\bar{\bm x}_{\text{RAD}}$ are used as the input to the cross-attention module.

% The standardized features, denoted as $\bar{X_r}$, are derived by subtracting the mean and scaling to unit variance. Subsequently, $\bar{X_r}$ are used as the input to the cross-attention module.

%The standardized features, $\bar{X_r}$, are performed by removing the mean and scaling to unit variance with Python package, sklearn \verb|standardscaler|\footnote{https://scikit-learn.org/stable/}. Then $\bar{X_r}$ serves as the input to the cross-attention module. %The process of extracting radiomic features can be defined as below:

% \begin{equation}\label{pyradiomic}
% \begin{split}
% X_r &= PyRadiomics(X^{C \times H \times W}, X^{H \times W}_{roi}) \\
% \bar{X_r} &= StandardScaler(X_r)
% \end{split}
% \end{equation}

% \begin{equation}\label{pyradiomic}
% \bar{X_r} = \verb|standardscaler|(\verb|pyradiomics|(X^{C \times H \times W}, X^{H \times W}_{roi}))
% \end{equation}
% where $\bar{X_r}$ represents the output features after radiomic feature extraction process and serves as the input to the cross-attention module.

\subsection{Feature Fusion}

% \hl{TODO: Watch out your notation, should be consistent. scalar italic e.g. $x$, vector bold $\bm x$, matrix upper case $W$ or $\text{W}$ but keep it consistent. I started to modify but need to be completed. Maybe used XA for Cross-attention subscript?}

We pass features to a cross-attention module to perform fusion operation, see Fig.~\ref{modules}(b). The cross-attention module in the model has an embedding size of 512 and is linked to a multi-layer perceptron (MLP) module. The MLP module consists of two fully connected layers with a hidden feature size of 512 and the $\ReLU$ activation layer. Firstly, input feature $\bar{\bm x}_{\text{RAD}}$ is linearly projected to queries ${\bm q}=W_\text{Q} \bar{\bm x}_{\text{RAD}} + b_\text{Q}$ $({\bm q} \in \mathbb R^{d_e})$, and feature ${\bm x}_\text{DL}$ is linearly projected to keys ${\bm k}=W_\text{K} {\bm x}_\text{DL} + b_\text{K}$ $({\bm k} \in \mathbb R^{d_e})$, and values ${\bm v}=W_\text{V} {\bm x}_\text{DL} + b_\text{V}$ $({\bm v} \in \mathbb R^{d_e})$. 
%The projection process can be defined as follows:
% \begin{equation}
% \begin{split}
% Q &= Linear(\bar{X_r}, d\_emb) \\
% K &= Linear(X_f, d\_emb) \\
% V &= Linear(X_f, d\_emb) 
% \end{split}
% \end{equation}
$d_e$ represents the dimension of hidden embeddings. In our framework, $d_e$ is set to 512. Next, the scaled dot-product attention is employed in the cross-attention module. Mathematically, it can be expressed as:

\begin{equation}
X_\text{XA} = \operatorname{softmax}\left(\frac{{\bm q} {\bm k}^{\top}}{\sqrt{d_{\bm k}}}\right) {\bm v} = {A}_\text{XA} {\bm v}
\end{equation}
where ${A}_\text{XA} \in \mathbb{R}^{d_{\bm q} \times d_{\bm k}}$ is the attention weight matrix, and $d_{\bm k}$ is the feature dimension of ${\bm k}$. The attention score $({\bm q} {\bm k}^{\top} \in \mathbb{R}^{d_{\bm q} \times d_{\bm k}})$ is computed between each query and all keys. Normalization is done across the key dimension to obtain attention weights $\text{A}_\text{XA} = \operatorname{softmax}(\cdot)$.
The MLP module processes the cross-attention module's output to predict GA (denoted as $\hat{y} \in \mathbb{R}$) as follows:

\begin{align}
\bm{y}_\text{XA} &= \ReLU( W_\text{XA} \cdot \vecop(X_\text{XA}) + b_\text{XA}\;) \\
\hat{y} &= W_\text{MLP} \cdot \bm{y}_\text{XA} + b_\text{MLP}
\end{align}
where $\vecop(\cdot)$ is the flatten function, and $\bm{y}_\text{XA} \in \mathbb{R}^{512}$ is a feature vector.

\subsection{Model Training and Loss Functions}

\def\angled#1{{\langle #1 \rangle}}

We employ a fine-tuning strategy for our framework, initializing the DL models with pre-trained weights from ImageNet \cite{Deng:2009}. %{\color{blue} Only the trainable components in our proposed framework are fine-tuned, see Fig.~\ref{fusion_network}.} 
We optimize the likelihood of the $\hat{y}$ regression output and therefore use the least squares for the overall training loss: $\mathcal{L} = \sum_{i=1}^M\left(y^\angled{i}-\hat{y}^\angled{i}\right)^2$, 
%
% \hl{Note: As mentioned, taking $\sqrt$ and dividing by $N$ does not change anything in terms of optimization, just wasting computation resources. Your loss should be the least squares. RMSE is used as performance measure. Also I generally prefer to use $M$ for the cardinality of the training set.} 
%
% \begin{equation}
% % \mathcal{L}(\hat{y},y_i) = \sqrt{MSE(\hat{y},y_{gt})}
% % \mathrm{RMSD}=\sqrt{\frac{\sum_{i=1}^N\left(x_i-\hat{x}_i\right)^2}{N}}
% \mathcal{L} = \sum_{i=1}^M\left(y^\angled{i}-\hat{y}^\angled{i}\right)^2
% \end{equation}
where $\hat{y}^\angled{i}$ is the model predicted GA for the $i^\text{th}$ training input and $y^\angled{i}$ is the computed GA as described in Section~\ref{data_desc}. $M$ is the size of the training set. The model with the best performance is saved for testing based on their predictive accuracy during the model training phase.

\textbf{Machine Learning.} In our extensive analysis, we build traditional ML models for predicting GA using radiomic features exclusively. These models include %Linear Regression, 
Support Vector Machine (SVM), K-Nearest Neighbors (KNN), Random Forest (RF), AdaBoost, Ridge and Gradient Boosting Regression (GBR). Additionally, we perform feature selection, including recursive feature elimination (REF) and least absolute shrinkage and selection operator (LASSO), to improve the performance of ML models.

\section{Experiments and Results}
\subsection{Datasets}
\label{data_desc}
We utilize two public datasets in this study: the Spanish trans-thalamic dataset (\textbf{ES-TT}) originates two centers in Spain \cite{Xavier:2020}; the \textbf{HC18} dataset is sourced from a database in Netherlands \cite{Heuvel:2018_b}. 
% {\color{blue} 
All scans were acquired from healthy singleton pregnancies. 
% HC18 data was collected using two ultrasound machines from a single center, while ESTT data was obtained from six machines across two different centers.
% } 
% of the Department of Obstetrics of the Radboud University Medical Center, Nijmegen, the Netherlands \cite{Heuvel:2018,Heuvel:2018_b}. 
The variability of the fetal head across various pregnancy trimesters poses a challenge in estimating GA. %{\color{blue}We combine two datasets into one and use a 70:30 ratio to split the data into training and testing sets.} 
We use a 70:30 ratio to split the data into training and testing sets. A subset of 100 images from the test set is randomly selected for validation purposes.

\textbf{ES-TT.} The images in this dataset are captured when pregnant women are in their second and third trimesters undergoing routine examinations 
% omitting instances of multiple pregnancies, congenital malformations, and aneuploidies
\cite{Xavier:2020,Alzubaidi:2023}. 
% The images in this dataset are captured between October 2018 and April 2019 as part of standard clinical practices. The study includes pregnant women in their second and third trimesters who are undergoing routine examinations, omitting instances of multiple pregnancies, congenital malformations, and aneuploidies \cite{Xavier:2020,Alzubaidi:2023}. The total number of images is 1552. 
The training set comprises 1086 images from 704 patients, while the test set includes 466 images from 391 patients. Students, experienced physicians, and radiologists annotated the ROI of the fetal head in all images. A file detailing the pixel calibration, $p_\text{size}$ (in mm), for each image is also provided.

% Train: 1086 images (704 patients).
% Test: 466 images (391 patients).

\textbf{HC18.} %Ultrasound images are obtained from 551 pregnant women who receive a standard ultrasound screening examination. 
This dataset includes fetuses without any growth abnormalities. The images are captured by experienced sonographers %utilizing the Voluson E8 or Voluson 730 ultrasound machines 
\cite{Heuvel:2018_b}. In each image, the sonographer draws an ellipse as ROI to best fit the circumference of the fetal head. We utilize 799 annotated images that are publicly accessible. The training set has 697 images from 603 patients, while the test set has 300 images from 279 patients. The pixel calibration, $p_\text{size}$ (in mm), is provided for each image.

% The reference GA is determined with a CRL measurement between 20 mm ($8+4$ weeks) and 68 mm ($12+6$ weeks) \cite{Heuvel:2018_b}. 

% Train: 697 images (603 patients).
% Test: 300 images (279 patients).

%To decrease computational demands, images and masks are resized to a lower resolution of $224 \times 224$ for training the Vision Transformer model and to $256 \times 256$ for training other DL models. Images and masks are also normalized using mean $[0.485,0.456,0.406]$ and standard deviation (std) $[0.229,0.224,0.225]$.

\textbf{Formula Calculated GA.} The ES-TT dataset provides the pixel size $p_\text{size}$ for every ultrasound image in millimeters, while the HC18 dataset presents the HC values in millimeters corresponding to each image. To obtain the GA for all images, we calculate the number of pixels, denoted as $p_\text{num}$, along the edge of the respective ROI in the input image $X_\text{ROI}$. Subsequently, we compute the corresponding HC in millimeters (mm) using the formula: $HC = p_\text{num} \times p_\text{size}$.
We obtain the GA ($y^\angled{i}$) using the equation derived from previous research \cite{Papageorghiou:2016}: 

\noindent $GA=\exp \left[0.05970 \times\left(\log_e(\text{HC})\right)^2+0.000000006409 \times(\text{HC})^3+3.3258\right]$.

% \subsection{Evaluation Metrics}
% Two evaluation metrics are used to assess the accuracy of predicting GA: Mean Absolute Error (MAE) and Root Mean Square Error (RMSE). In evaluation of ML models, we additionally include the statistical metric R-Squared ($R^2$), also called the coefficient of determination, to evaluate how well a regression model fits the data.

% COMMENT: Already described
% The DL model $f_{\theta}$ is followed by a fully connected layer with an output dimension of 512. 
\subsection{Implementation Details} 
The input image $X_\text{US}$ and mask $X_{\text{ROI}}$ are resized to the same resolution, $256 \times 256$. %To decrease computational demands, $X_\text{US}$ and $X_{\text{ROI}}$ are normalized using mean $[0.49,0.46,0.41]$ and standard deviation $[0.229,0.224,0.225]$. 
%{\color{blue} $X_\text{US}$ is normalized following the same process from \cite{He:2016}}. 
The Adam optimizer is utilized with an initial learning rate of $10^{-5}$ following a weight decay of $10^{-6}$. The batch size is fixed at 8, and the epoch is set to 60. Our framework is implemented using Pytorch library and trained on a single RTX 4090 GPU with 24 GB memory.

\textbf{Data Augmentation.} During model training, we apply common data augmentation techniques on the images within the training set. Specifically, these augmentation techniques randomly rotate the images within range $(-15^\circ, 15^\circ)$ and randomly flip them horizontally with a probability of 50\%.

\subsection{Results}

\begin{table}[tb]
\parbox{.5\linewidth}{
\centering
\caption{Test results of GA estimation by using DL models with concatenation and cross-attention mechanisms. C: Concatenation. XA: Cross-attention.}\label{test_results}
% \begin{tabular}{lcc|ll|c}
% \hline
% \multirow{2}{*}{DL Model} & \multirow{2}{*}{C} & \multirow{2}{*}{XA} & Avg. & Avg. & P \\
% &  &  & RMSE & MAE & value\\
% \hline
% ResNet18 & \xmark & \xmark & 13.60 & 9.87 & \multirow{2}{*}{<} \\
% EfficientNet V2 & \xmark & \xmark & 12.80 & 9.27 & \multirow{3}{*}{0.001} \\ % \cite{Wang:2025} 
% MaxViT & \xmark & \xmark & 14.78 & 10.72 & \\
% SwinTransformer & \xmark & \xmark & 12.85 & 9.44 & \\
% ConvNeXt & \xmark & \xmark & 12.26 & 8.59 & \\
% \hline
% ResNet18 & \cmark & \xmark & 14.01 & 10.23 & \multirow{2}{*}{<} \\
% EfficientNet V2 & \cmark & \xmark & 11.94 & 8.58 & \multirow{3}{*}{0.001} \\
% MaxViT & \cmark & \xmark & 13.26 & 9.56 & \\
% SwinTransformer & \cmark & \xmark & 12.36 & 8.61 & \\
% ConvNeXt & \cmark & \xmark & 12.32 & 8.72 & \\
% \hline
% ResNet18 & \xmark & \cmark & 14.15 & 10.35 & \multirow{5}{*}{-} \\
% EfficientNet V2 & \xmark & \cmark & 12.19 & 8.65 & \\
% MaxViT & \xmark & \cmark & 12.12 & 8.66 & \\
% SwinTransformer & \xmark & \cmark & 12.01 & 8.31 & \\
% \textbf{ConvNeXt} & \xmark & \cmark & \textbf{11.52} & \textbf{8.04} & \\
% \hline
% \end{tabular}
\setlength{\tabcolsep}{2.5pt}
\begin{tabular}{|lcc|c|c|}
\hline
DL Model & C & XA & MAE $\downarrow$ & P-value\\
\hline
ResNet18~\cite{Lee:2023} & \xmark & \xmark & 9.9$\pm$0.02 & \\
EfficientNet & \xmark & \xmark & 9.3$\pm$0.00 & \multirow{3}{*}{<$10^{-3}$ } \\ % \cite{Wang:2025} 
MaxViT & \xmark & \xmark & 10.7$\pm$0.02 & \\
SwinTrans & \xmark & \xmark & 9.4$\pm$0.01 & \\
ConvNeXt & \xmark & \xmark & 8.6$\pm$0.01 & \\
\hline
ResNet18 & \cmark & \xmark & 10.2$\pm$0.01 & \\
EfficientNet & \cmark & \xmark & 8.6$\pm$0.02 & \multirow{3}{*}{<$10^{-3}$ } \\
MaxViT & \cmark & \xmark & 9.6$\pm$0.04 & \\
SwinTrans & \cmark & \xmark & 8.6$\pm$0.01 & \\
ConvNeXt & \cmark & \xmark & 8.7$\pm$0.01 & \\
\hline
ResNet18 & \xmark & \cmark & 10.4$\pm$0.01 & \multirow{5}{*}{--} \\
EfficientNet & \xmark & \cmark & 8.6$\pm$0.00 & \\
MaxViT & \xmark & \cmark & 8.7$\pm$0.00 & \\
SwinTrans & \xmark & \cmark & 8.3$\pm$0.00 & \\
\textbf{ConvNeXt} & \xmark & \cmark & \textbf{8.0$\pm$0.00} & \\
\hline
\end{tabular}
}
\hfill
\parbox{.47\linewidth}{
\centering
\caption{Test results of GA estimation using ML models by adopting LASSO and RFE techniques to select radiomic features. SVM: Support Vector Machine. KNN: K-Nearest Neighbors. RF: Random Forest. GBR: Gradient Boosting Regression.}\label{ml_models}
% \begin{tabular}{l|lll}
% \hline
% \multirow{2}{*}{Model} & Avg. & Avg. & \multirow{2}{*}{$R^2$} \\
% & RMSE & MAE & \\
% \hline
% SVM (LASSO) & 34.97 & 25.18 & 0.62 \\ 
% KNN (LASSO) & 35.56 & 25.20 & 0.61 \\ 
% AdaBoost (LASSO) & 35.72 & 26.78 & 0.60 \\ 
% Ridge (LASSO) & 34.04 & 25.74 & 0.64 \\ 
% RF (LASSO) & 31.55 & 21.65 & 0.69 \\
% GBR (LASSO) & 31.65 & 21.84 & 0.69 \\
% \hline
% SVM (RFE) & 34.85 & 24.89 & 0.62 \\ 
% KNN (RFE) & 34.75 & 24.52 & 0.62 \\ 
% AdaBoost (RFE) & 35.55 & 26.51 & 0.61 \\ 
% Ridge (RFE) & 34.06 & 25.78 & 0.64 \\ 
% RF (RFE) & 31.54 & 21.38 & 0.69 \\
% \textbf{GBR (RFE)} & 31.02 & \textbf{21.29} & \textbf{0.70} \\
% \hline
% \end{tabular}
\setlength{\tabcolsep}{4pt}
\begin{tabular}{|l|c|c|}
\hline
ML Model & MAE $\downarrow$ & $R^2$ $\uparrow$ \\
\hline
SVM (LASSO) & 25.2$\pm$24.3 & 0.62 \\ 
KNN (LASSO) & 25.2$\pm$25.1 & 0.61 \\ 
AdaBoost (LASSO) & 26.8$\pm$23.6 & 0.60 \\ 
Ridge (LASSO) & 25.7$\pm$22.3 & 0.64 \\ 
RF (LASSO) & 21.7$\pm$23.0 & 0.69 \\
GBR (LASSO) & 21.8$\pm$22.9 & 0.69 \\
\hline
SVM (RFE) & 24.9$\pm$24.4 & 0.62 \\ 
KNN (RFE) & 24.5$\pm$24.6 & 0.62 \\ 
AdaBoost (RFE) & 26.5$\pm$23.7 & 0.61 \\ 
Ridge (RFE) & 25.8$\pm$22.3 & 0.64 \\ 
RF (RFE) & 21.4$\pm$23.1 & 0.69 \\
GBR (RFE) & 21.3$\pm$22.6 & 0.70 \\
\hline
\end{tabular}
}
\end{table}

% \subsubsection{Estimation of GA}
To assess the accuracy of predicting GA, we use Mean Absolute Error (MAE) and Root Mean Square Error (RMSE) as the evaluation metrics. We find that the correlation coefficient between MAE and RMSE is larger than 0.998, therefore, we only report the  MAE in this paper. %{\color{blue} We evaluate our method by comparing it with Baseline utilizing only images and Concatenation utilizing concatenated image and radiomic features.}
% One common evaluation metric is used to assess the accuracy of predicting GA: Mean Absolute Error (MAE). 

From Table~\ref{test_results}, we observe several key findings. Our proposed framework has achieved the best result, %with an average RMSE of 11.52 and 
with an average MAE of 8.0 (days). 
Quantitative analysis shows a significant improvement in accurate GA estimation with the cross-attention module compared to both the concatenation module ($p < 0.001$) and the baseline model ($p < 0.001$). Our method enhances the baseline method, ResNet18~\cite{Lee:2023}, by 1.9 days in accurately estimating GA.
Despite ResNet18 and EfficientNet models exhibiting higher MAE values when utilizing the cross-attention module than the concatenation module, the overall accuracy across all architectures is significantly improved with the former ($p < 0.001$), showcasing the superiority of the cross-attention mechanism. It is notable that ConvNeXt, an efficient model with 3.7M parameters, demonstrates superior performance compared to four other DL-based feature extractors, as evidenced in Table~\ref{test_results}.

Table~\ref{ml_models} shows the test results of various ML algorithms without using the radiomic features. The statistical metric $R$-squared ($R^2$), also known as the coefficient of determination, is employed to assess the adequacy of a model in fitting the data. The $R^2$ value suggests that GBR (RFE) is a more effective model for capturing the radiomic features than other ML algorithms. However, all ML algorithms exhibit significantly higher MAE values in GA estimation compared to our proposed method, highlighting the superiority of our method over traditional ML algorithms.

% \begin{table}
% \centering
% \caption{Test results for estimating GA using ML models by employing two techniques to select radiomic features. SVM: Support Vector Machine. KNN: K-Nearest Neighbors. GBR: Gradient Boosting Regression.}\label{ml_models}
% \begin{tabular}{lcc|lll}
% \hline
% \multirow{2}{*}{Model} & \multirow{2}{*}{RFE} & \multirow{2}{*}{LASSO} & Avg. & Avg. & \multirow{2}{*}{$R^2$} \\
% &  &  & RMSE & MAE & \\
% \hline
% SVM Tree & \xmark & \cmark & 34.97 & 25.18 & 0.62 \\ 
% % Decision Tree &  & \cmark & 37.76 & 24.51 & 0.56 \\ 
% KNN Tree & \xmark & \cmark & 35.56 & 25.20 & 0.61 \\ 
% AdaBoost & \xmark & \cmark & 35.72 & 26.78 & 0.60 \\ 
% Ridge Tree & \xmark & \cmark & 34.04 & 25.74 & 0.64 \\ 
% Random Forest & \xmark & \cmark & 31.55 & 21.65 & 0.69 \\
% GBR & \xmark & \cmark & 31.65 & 21.84 & 0.69 \\
% \hline
% SVM Tree & \cmark & \xmark & 34.85 & 24.89 & 0.62 \\ 
% % Decision Tree & \cmark &  & 37.21 & 24.42 & 0.57 \\ 
% KNN Tree & \cmark & \xmark & 34.75 & 24.52 & 0.62 \\ 
% AdaBoost & \cmark & \xmark & 35.55 & 26.51 & 0.61 \\ 
% Ridge Tree & \cmark & \xmark & 34.06 & 25.78 & 0.64 \\ 
% Random Forest & \cmark & \xmark & 31.54 & 21.38 & 0.69 \\
% \textbf{GBR} & \cmark & \xmark & \textbf{31.02} & \textbf{21.29} & \textbf{0.70} \\
% \hline
% \end{tabular}
% \end{table}

\begin{figure}[htbp]
\centering
\includegraphics[width=\textwidth]{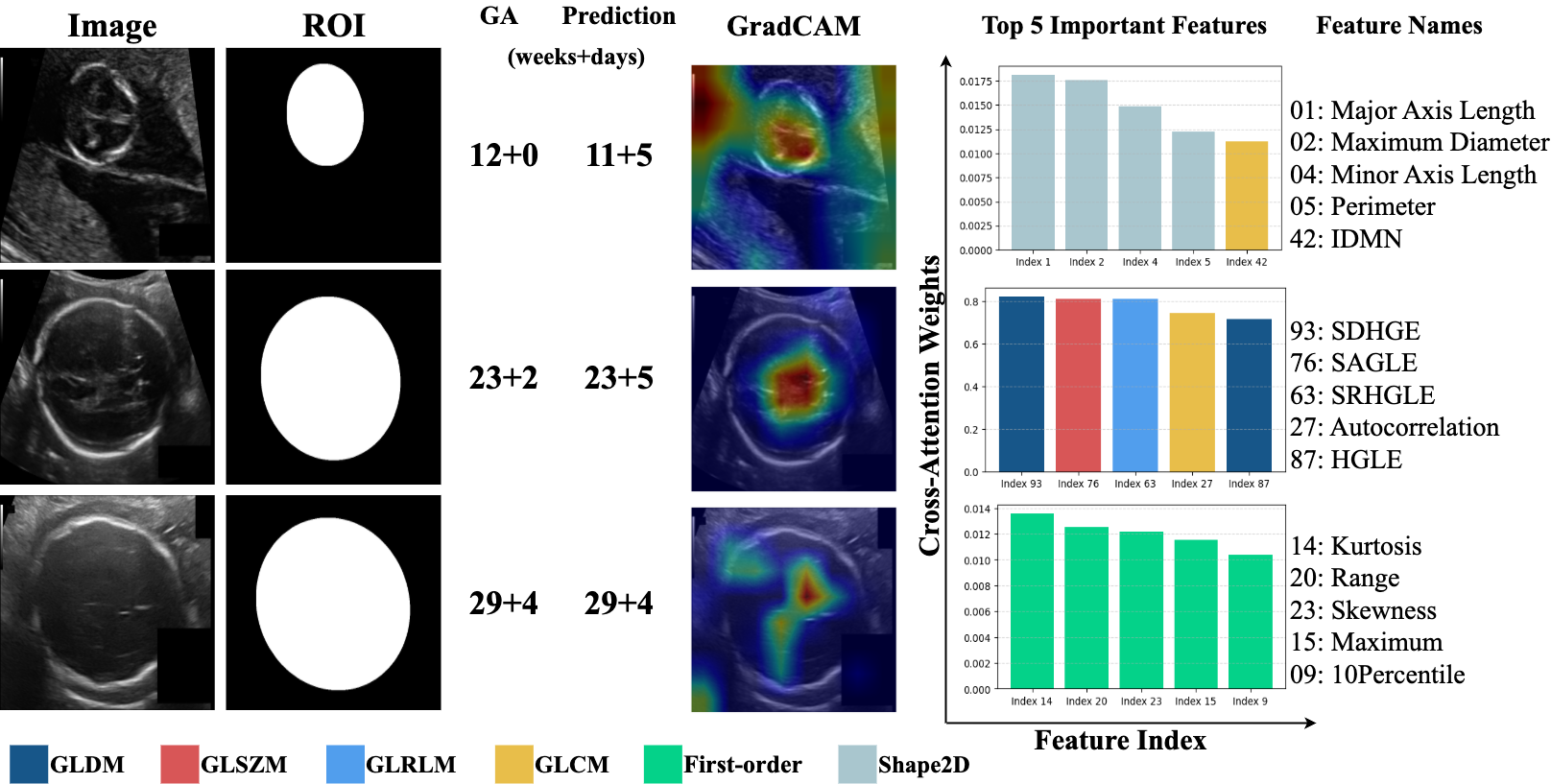}
\caption{The visualization of the interpretability of our method. IDMN: Inverse Difference Moment Normalized. SDHGE: Small Dependence High Gray Level Emphasis. SAGLE: Small Area High Gray Level Emphasis. SRHGLE: Short Run High Gray Level Emphasis. HGLE: High Gray Level Emphasis.} \label{att_weight}
\end{figure}

% {\color{blue} 
For better understanding, we visualize the GradCAM-based explanations \cite{Selvaraju:2017} and cross-attention-based feature attribution of three test cases from the $1^\text{st}$, $2^\text{nd}$, and $3^\text{rd}$ trimesters in Fig.~\ref{att_weight}. 
The GradCAM visualizations of our ConvNeXt layer show that the image model focuses on the brain tissue in the $2^\text{nd}$ and $3^\text{rd}$ samples, whereas in the $1^\text{st}$ sample, it concentrates on the entire ROI along with some background pixels. 
The bar charts are employed to visualize the five most influential radiomic features for GA estimation, as determined by cross-attention weights. In the $1^\text{st}$ and $3^\text{rd}$ samples, Shape2D and First-order features are the primary contributors to the estimation. In contrast, in the $2^\text{nd}$ sample, a diverse set of features such as GLDM, GLSZM, GLRLM, and GLCM play a significant role in the prediction.
% }

% For better understanding, we visualize the cross-attention maps of three test cases from the $1^\text{st}$, $2^\text{nd}$, and $3^\text{rd}$ trimesters in Fig.~\ref{att_weight}. It can be observed that the cross-attention module plays a crucial role by offering an overview for identifying distinctive features and exploiting synergistic interactions among them. For example, the radiomic features and deep representations highlight similar patterns in fetal data from $2^\text{nd}$ and $3^\text{rd}$ trimesters. On the other hand, the cross-attention weights in the first trimester fetal data demonstrate the DL model's ability to focus on two radiomic features: Inverse Difference and Inverse Difference Moment. Attention weight visualization an interpretable approach to understanding the contribution of each individual radiomic feature on the overall significance of GA estimation.

\begin{table}[tb]
\centering
\caption{Ablation study on the different feature fusion modules and pre-trained weights. C: Concatenation. XA: Cross-attention.}\label{ablation}
% \begin{tabular}{lcccc|ll}
% \hline
% Model & Image & Pretrain & Concat & Cross-attention & RMSE $\downarrow$ & MAE $\downarrow$ \\
% \hline
% ConvNeXt & \cmark & \xmark & \xmark & \xmark & 35.84 & 28.24 \\
% ConvNeXt & \cmark & \cmark & \xmark & \xmark & 12.26 & 8.59 \\
% ConvNeXt $+$ CONCAT & \cmark & \cmark  & \cmark & \xmark & 12.32 & 8.72 \\
% \textbf{ConvNeXt $+$ XA (Ours)} & \cmark & \cmark & \xmark & \cmark & \textbf{11.52} & \textbf{8.04} \\
% \hline
% \end{tabular}
\setlength{\tabcolsep}{4pt}
\begin{tabular}{|lcccc|c|}
\hline
Model & Image & Pre-trained & C & XA & MAE $\downarrow$ \\
\hline
ConvNeXt & \cmark & \xmark & \xmark & \xmark & 28.9$\pm$0.04 \\ %4+2 \\
ConvNeXt & \cmark & \cmark & \xmark & \xmark & 8.6$\pm$0.01 \\ %1+2 \\
ConvNeXt $+$ C & \cmark & \cmark  & \cmark & \xmark & 8.7$\pm$0.01 \\ %1+2 \\
\textbf{ConvNeXt $+$ XA (Ours)} & \cmark & \cmark & \xmark & \cmark & \textbf{8.0$\pm$0.00} \\ %1+1 \\
\hline
\end{tabular}
\end{table}

\textbf{Ablation Study.} We investigate the effectiveness of three primary components in our framework: fine-tuning strategy, feature fusion strategy, and cross-attention module. We first set up a baseline framework without using these components. Table~\ref{ablation} illustrates the significance of fine-tuning the framework with a pre-trained DL model in enhancing the accuracy of predicting GA. The feature fusion strategy using the concatenation method demonstrates minimal significance in our framework. However, integrating cross-attention into our framework substantially enhances the accuracy of predicting GA. This indicates the cross-attention module in facilitating a more effective fusion of deep representations and radiomic features.

\section{Conclusion}

This study introduces a novel feature fusion framework that combines image and radiomic features to estimate fetal GA from ultrasound images. This framework can be developed into an interpretable tool for clinical use.
% {\color{blue} 
We validate our framework on two fetal head ultrasound image datasets and achieve better results than those achieved by existing image-based DL and radiomic-based ML methods. Moreover, our framework can readily be utilized for analyzing additional fetal anatomical structures in ultrasound, such as the fetal abdomen and femur.
% } %Moreover, our method is a general framework and can be expanded to encompass additional fetal anatomical structures in ultrasound datasets, including the fetal abdomen and femur.
Hence, our proposed framework is a valuable algorithm that collaborates with a segmentation model to estimate GA, thereby benefiting patients and clinicians in clinical practice. Our future work includes enhancing the performance of GA estimations by utilizing a combination of multiple metrics obtained from various standard ultrasound planes across diverse cohorts. These experiments will be validated using larger datasets.

\begin{credits}
\subsubsection{\ackname} This work was funded by Taighde \'{E}ireann – Research Ireland through the Research Ireland Centre for Research Training in Machine Learning \\
(18/CRT/6183).

\subsubsection{\discintname}
The authors have no competing interests to declare that are relevant to the content of this article. 
\end{credits}

\bibliographystyle{splncs04}
\bibliography{mybibliography}

\end{document}